\documentclass[a4paper,11pt]{article}
\usepackage{pos}

\title{Neutrino flavor evolution in dense environments and the r-process}

\author*{Maria Cristina Volpe}

\affiliation{Astro-Particule et Cosmologie (APC), CNRS UMR 7164, Universit\'e Denis Diderot,\\ 10, rue Alice Domon et L\'eonie Duquet, 75205 Paris Cedex 13, France}

\emailAdd{volpe@apc.univ-paris7.fr}

\abstract{In dense environments, standard and non-standard neutrino interactions with the background particles trigger a variety of flavor mechanisms, which can impact r-process nucleosynthetic abundances. Future observations of a(n) (extra)galactic supernova will tell us about properties of neutrinos and of the astrophysical source that produce them. The upcoming measurement of the diffuse supernova neutrino background constitute a unique source of information. We highlight some recent developments.}

\FullConference{%
  *** Particles and Nuclei International Conference - PANIC2021 ***\\
  *** 5 - 10 September, 2021 ***\\
  *** Online ***
}

\begin{document}
\maketitle

\section{Introduction}
\noindent
Confirmation that core-collapse supernovae emit 99 $\%$  of a few $10^{53}$ erg as neutrinos was given by the measurement of the 24 SN1987A $\nu$ events in Kamiokande \cite{Hirata:1987hu}, IMB \cite{Bionta:1987qt} and Baksan \cite{Alekseev:1988gp}. Core-collapse supernovae (CC SNe) and binary neutron star mergers (BNS) are candidate sites for the synthesis of elements heavier than iron through the rapid neutron capture process ($r$-process). Weak processes involving neutrinos play an important role in these sites, for the explosion dynamics, for future observations and for r-process nucleosynthesis. 
Evidence for the production of r-process elements in BNS was provided by the kilonova observed with GW170817 \cite{LIGOScientific:2017vwq}. Gravitational wave measurements will furnish, among others, a precise measurement of the BNS rate, crucial for unravelling the sites for the $r$-process, a longstanding open issue.  

Numerous theoretical investigations clearly show that  flavor evolution impacts the $r$-process. Challenging self-consistent multidimensional calculations of the matter composition in neutrino driven winds and neutrino flavor evolution are still missing, A reference phenomenon, for flavor evolution studies, is the Mikheev-Smirnov-Wolfenstein (MSW) mechanism \cite{Wolfenstein:1977ue,Mikheev:1986wj}, due to neutrino-matter interactions, a phenomenon responsible for high energy solar $^8$B neutrinos to 1/3.  In dense environments, neutrinos change flavor due to neutrino self-interactions (see e.g. \cite{Duan:2010bg,Mirizzi:2015eza} for a review), shock waves and turbulence. Pinning down the conditions, their impact and their description has triggered intense investigations since more than a decade.  Moreover, efforts have been made to extend the current theoretical description of neutrino flavor evolution in dense media and questioned the (widely employed) mean-field approximation (see e.g. \cite{Volpe:2015rla,Patwardhan:2021rej}). Here we highlight some MSW-like  mechanisms found in BNS based on the mean-field neutrino evolution equations. 

 \section{Flavor evolution from standard and non-standard interactions in BNS} 
 \noindent
 BNS produce large amount of low energy neutrinos with luminosities and average energies comparable to CC SNe, but with an excess of $\bar{\nu}_e$ over $\nu_e$, due to the matter neutron richness and the geometry of the neutrino emission. Moreover the $\nu_{\mu}, \nu_{\tau}$ fluxes can be much smaller than $\nu_e$ ones. The flux predictions present sizeable variations depending on simulations (see Table VII of Ref.\cite{Frensel:2016fge}).  
 
 The $\bar{\nu}_e$ excess in BNS can produce a cancellation between the matter and the neutrino self-interactions contributions to the neutrino propagation Hamiltonian. This cancellation can trigger  the {\it matter-neutrino resonance} (MNR) and adiabatic conversion of  $\nu_e$ (standard) and $\bar{\nu}_e$ (symmetric MNR), influencing $r$-process nucleosynthetic abundances \cite{Malkus:2015mda}. The flavor content depends on the initial conditions for the neutrino emission and the trajectories followed by the neutrinos, modifying $\nu_e$ and $\bar{\nu}_e$ rates on neutrons and protons by several tens of percent depending on the neutrino trajectories \cite{Frensel:2016fge}.

The MNR underlying mechanisms is a multiple MSW effect, as shown by Ref.\cite{Wu:2015fga} with a schematic model, and by Ref.\cite{Chatelain:2016xva} based on detailed BNS simulations. By using a perturbative argument, one can show that the $\nu$ self-interaction adjusts itself to the matter term multiple times, due to the non-linear feedback. On the contrary, non-linear feedback does not operate for {\it helicity coherence}  that arises because of wrong helicity contributions to the evolution equations, due to the neutrino mass \cite{Chatelain:2016xva}. 

Constraints on non-standard interactions (NSI) come from solar, oscillation experiments, scattering measurements \cite{Farzan:2017xzy} and coherent neutrino-nucleus scattering \cite{COHERENT:2017ipa}. NSI can trigger new flavor conversion mechanisms in dense environments, such as the MNR in core-collapse supernovae \cite{Stapleford:2016jgz} or
the I-resonance, triggered by a cancellation of standard and non-standard matter terms \cite{Esteban-Pretel:2009jqw}.

The first investigation of NSI effects in BNS has uncovered that
the I-resonance can also be triggered in presence of sizeable $\nu$ self-interactions as a synchronized MSW effect. The study has also shown that, as in the supernova context, a complex pattern of flavor conversion mechanisms takes place along different trajectories, even for very small values of the NSI couplings (well below current bounds). This can influence significantly the electron fraction and nucleosynthetic abundances  \cite{Chatelain:2017yxx}.

Moreover, in dense environments, very short scale conversions, termed {\it fast} modes, can occur when more realistic emission at the neutrinosphere is considered  \cite{Sawyer:2015dsa}, with crossings between the $\nu_e$ and $\bar{\nu}_e$ angular distributions (see \cite{Tamborra:2020cul} for a review). 
Such modes, that do not appear to produce flavor equilibration (in a two neutrino beam schematic model) \cite{Abbar:2018beu},  can take place in detailed multidimensional simulations (when the ratio of $\nu_e$ and $\bar{\nu}_e$ number densities is close to one)  \cite{Abbar:2018shq}. This result was confirmed by analysis of multidimensional simulations from different groups. 

 \section{Future supernova neutrino observations}
 \noindent
The Supernova Early Warning System (SNEWS)  is awaiting for the next (extra)galactic supernova \cite{SNEWS:2020tbu} that will bring invaluable information of the supernova explosion mechanism, the supernova location, on neutrino properties, non-standard particles, NSI and on the neutron star. 

The gravitational binding energy can be determined with $11 \%$ precision in Super-Kamiokande and $3 \%$ in Hyper-Kamiokande, with likelihood analysis of the neutrino fluences (MSW effect included) and combining inverse beta-decay and elastic scattering. The $\bar{\nu}_e, \nu_{\mu}, \nu_{\tau}$ flux parameters can be precisely reconstructed (except the pinching parameters). Based on the EOS of neutron stars, the compactness or mass-radius relation can be inferred with rather good precision \cite{GalloRosso:2017hbp,GalloRosso:2017mdz}.

Complementary to future supernova neutrino observations is the measurement of the diffuse supernova background from past supernovae (DSNB). 
The DSNB is sensitive to redshifted neutrino spectra and flavor mechanisms, the core-collapse supernova rate, the debated fraction of failed supernovae \cite{Kresse:2020nto,Horiuchi:2020jnc}and $\nu$ properties, such as neutrino decay \cite{DeGouvea:2020ang}. As for flavor phenomena, beyond the MSW, shock waves (including phase effects) and $\nu$ self-interactions impact the DSNB fluxes and rates. Shock wave effects can change the latter by about 10-20 $\%$ (depending on the $\nu$ mass ordering) and can reduce the sensitivity to neutrino self-interactions \cite{Galais:2009wi}. The DSNB rates are also sensitive to the shock revival time \cite{Nakazato:2013maa}.

The sensitivity of the combined analysis (90 $\%$ C.L.) of SK-IV and previous phases, is on par with DSNB rates from four models, whereas more conservative predictions are lower about a factor of 4 \cite{Super-Kamiokande:2021acd}. The DSNB measurement is expected
with the running Super-Kamiokande (SK)+Gd and the upcoming JUNO and Hyper-Kamiokande experiments. Exciting observations are ahead of us.

\end{document}